\documentstyle[12pt,sprocl,phase1]{article}
\begin{document}

\

\vskip3cm

\Large
\centerline{\bf ~~~ Determination of Cosmological Parameters }

\bigskip
\bigskip
\bigskip
\bigskip

\centerline{~~~Wendy L. Freedman$^1$}
\normalsize 

\bigskip
\vskip3cm
\vskip3cm

\noindent
$^1$ Carnegie Observatories, 813 Santa Barbara St., Pasadena, CA 91101.

\bigskip
\bigskip

\noindent
{\it Invited Review given  at the Nobel Symposium, ``Particle  Physics
and  the Universe'',Haga Slott,  Sweden, August, 1998. To be published
by World Scientific Press}.

\vfill\eject

\large
\begin{center}
{\bf Determination of Cosmological Parameters }

\normalsize
{ Wendy L. Freedman}

{ Carnegie Observatories, Pasadena, CA}

\end{center}

\bigskip
\bigskip
\bigskip
\medskip

\large
\begin{center}
{\bf Abstract}
\end{center}

\normalsize

Rapid  progress has   been made  recently  toward  the measurement  of
cosmological parameters. Still, there are areas remaining where future
progress  will be relatively  slow and  difficult,  and where  further
attention is needed. In this review, the status of measurements of the
matter density ($\Omega_m$), the vacuum energy density or cosmological
constant ($\Omega_\Lambda$), the Hubble constant  (H$_0$), and ages of
the oldest   measured  objects (t$_0$) are   summarized.  Many recent,
independent dynamical measurements are  yielding  a low value for  the
matter density  ($\rm \Omega_m \sim$ 0.3).  New  evidence from type Ia
supernovae suggests  that $\Omega_\Lambda$    may be  non-zero.   Many
recent Hubble constant  measurements  appear to  be converging in  the
range of 65-75 km/sec/Mpc.  Eliminating systematic  errors lies at the
heart of   accurate measurements  for  all of  these parameters;  as a
result,  a wide range  of  cosmological  parameter space is  currently
still  open.  Fortunately,  the   prospects for  accurately  measuring
cosmological parameters continue to increase  and there is good reason
for optimism that success may shortly be forthcoming.

\bigskip
\bigskip

\medskip
\large
\begin{center}
{\bf Introduction and Brief Historical Overview}
\end{center}
\normalsize
\medskip

The recent success in the  measurement of cosmological parameters  can
be attributed to   a number    of    factors: an  abundance of     new
observations, new approaches,  and developments in detector technology
(with a corresponding  increase in the precision  of the data).  Given
these advances, it is tempting to conclude that we have now entered an
era of  precision  cosmology.  Of  course, whether  this is indeed the
case depends completely on the  extent to which systematic measurement
errors have  been minimized  or  eliminated.  In  this context,  it is
interesting to view the measurement of  cosmological parameters from a
historical perspective  as   described  briefly in the    next section
below. The present   review concentrates primarily  on  results of the
past  couple of years, following  on from a review  on a similar topic
given at the Texas Symposium in December, 1996 (Freedman 1997a).

The cosmological  parameters     discussed in this    review   are the
following: the matter density $\rm \Omega_m  = 8\pi G\rho_m / 3H_0^2$,
the Hubble parameter  H=${\dot{a}  \over a}$, (where  a is  the  scale
factor and  H$_0$, the Hubble constant,  is  the value at  the current
epoch),  the vacuum energy density    $\rm \Omega_\Lambda = \Lambda  /
3H_0^2$, and  the age of the universe,  t$_0$.  In Big Bang cosmology,
the Friedmann equation relates the density,  geometry and evolution of
the universe: $ \rm H^2 = { 8\pi G \rho_m \over 3} -  { k \over a^2} +
{ \Lambda \over  3}$, where the average mass  density  is specified by
$\rho_m$.  The curvature  term  is specified  by $\rm \Omega_k$  =  -k
/a$_0^2$H$_0^2$, and for  the case of a   flat universe (k =  0), $\rm
\Omega_m$  + $\rm  \Omega_\Lambda$ = 1.   A lower  limit  to the  age,
t$_0$, can be determined by dating the oldest objects in the Universe.
Or, alternatively,   given  an  independent knowledge   of   the other
cosmological parameters  (H$_0$, $\rm \Omega_m$, $\rm \Omega_\Lambda$,
and  $\Omega_k$),  a  dynamical  age  of  the   Universe  can also  be
determined by integrating the Friedmann  equation.

Before moving on to describe recent developments, it is interesting to
view how the  values for these parameters  have changed over time.  In
Figures 1a-b and 2a-b,  estimates of H$_0$, $\Omega_m$, $\Lambda$, and
t$_0$ are   shown as a  function of  time.   The historical discussion
below is not intended  to be comprehensive, but  rather to outline the
general trends  as shown in  Figures 1 and 2.   Each of the figures is
discussed in turn.

\medskip
{\bf  H$_0$} (Figure  1a): Over the   last half century,  the value of
H$_0$, the expansion rate of  the Universe, has remained a  well-known
source of disagreement  (for historical reviews  see, for example, van
den Bergh 1997; Rowan-Robinson 1985).  After Baade (1954) recalibrated
the Cepheid distance scale,   and Sandage (1958) recognized that   the
brightest stars   in galaxies were   ionized HII  regions, the  Hubble
constant decreased from its original value  of over 500, and fell into
the well-known range of a factor of two, loosely constrained, as shown
by the schematic lines drawn  in Figure 1a,  between about 50 and  100
km/sec/Mpc.  As indicated in  the figure and discussed  further below,
recent improvements have come about as a result of new instrumentation
and the availability of the Hubble Space Telescope (HST); in addition,
several  new  methods for measuring  distances  to  galaxies have been
developed (see also Livio,  Donahue \& Panagia 1997; Freedman  1997b).
Recently, there  has been some   convergence  in the  value of  H$_0$.
Although decreasing, the dominant errors in H$_0$ remain systematic in
nature.

\medskip
{$\bf  \Omega_m$}  (Figure 1b): Zwicky   (1933) provided evidence that
there was  possibly 10--100 times more  mass in the Coma  cluster than
contributed by the  luminous matter in galaxies.   However, it was not
until the 1970's  that the existence of dark  matter began to be taken
more seriously.  At  that time, evidence  began  to mount that  showed
rotation curves did  not fall off with  radius ({\it e.g.,} Rogstad \&
Shostak  1972; Rubin {\it et al.}   1978; and Bosma  1981) and that the
dynamical mass was increasing    with scale from that   of  individual
galaxies up  through groups of  galaxies ({\it e.g.}, Ostriker, Peebles
\& Yahil 1974).  A comprehensive historical review  of the dark matter
issue can be found in Trimble (1987).  By the 1970's, the evidence was
consistent with a total matter density of $\sim$10--20\% of a critical
density, ($\Omega$  = 1) universe.   With the development of inflation
in the    early 1980's (Guth  1981), tremendous    effort was aimed at
discovering both the  nature and the  amount  of dark matter.   In the
early 1990's (see, for example, the review by Dekel, Burstein \& White
1997),  a number  of studies  indicated   that we live   in a critical
density universe, and the  first results for high redshift  supernovae
(Perlmutter 1997) were also consistent with a high matter density.  As
described below, however, the new supernovae results, and a wide range
of other  studies are  consistent  (once again)   with a  lower matter
density  of  $\Omega_m  \sim$0.2   to 0.3.    The  overall trends   in
$\Omega_m$ with  time  are shown in   Figure 1b. The  solid and dotted
lines represent approximate upper and lower bounds only.

\medskip
{ $\bf  \Lambda$} (Figure  2a): The reader  is referred  to  excellent
overviews of the cosmological constant by Weinberg (1989) and Carroll,
Press \& Turner  (1992). Enthusiasm for  a non-zero value of $\Lambda$
has come and  gone several times  over this century.   For fun, I have
plotted  (in  arbitrary units),  the  ``market  value'' for  shares of
$\Lambda$ in  Figure 2a.   Here it can   be seen  that the  market for
$\Lambda$  has  been quite  volatile over time.   Shares for $\Lambda$
were  high when  Einstein  (1917) first  introduced this  cosmological
constant in  an  attempt to  allow  for a stable universe;  subsequent
work, for  example, by Friedmann  (1922), followed by the discovery of
the expansion  of the Universe  by Edwin Hubble,  led to the  crash of
$\Lambda$ (along with the rest of the stock market) in 1929!

The  value of H$_0$ measured by  Hubble (1929) implied a dynamical age
for the Universe of only $\sim$2 billion years.   This age was younger
than the geological dating estimates for the age  of the Earth, placed
then at  about 3.5 billion years.   This  discrepancy led to  an ``age
crisis'', and a renewal of interest in  $\Lambda$, that was eventually
solved by Baade's  recalibration  of the distance  scale  in 1954.   A
brief period of  activity  occured in the  $\Lambda$  market  with the
observation by Petrosian {\it et al.}   (1967), of an apparent peak at
z$\sim$2 in  the quasar distribution.  However,  as  more quasars were
observed, this feature also disappeared.

In general,  consumers have tended to been  wary of stock in $\Lambda$
due to the  difficulties of explaining  the current limits in conflict
by 120 orders of magnitude with the  predictions of the standard model
of particle physics ({\it  e.g.,} Weinberg 1989).  However,  recently,
as Figure  2a shows, the  low  observed matter  density, the recurring
conflict in ages  between some values   of H$_0$ and  globular cluster
ages,  and the observed   large-scale  distribution of galaxies   have
motivated  a  renewed interest in  $\Lambda$   ({\it e.g.},  Krauss \&
Turner   1995;  Ostriker  \& Steinhardt   1995).    Just  as for other
commodities, the consumer may need to be concerned about how inflation
is driving the  market.  It is  probably too early  to say if the bull
market  of the 1990's  is  over -  this is an  area  where, as  in the
world of the stock market, the experts disagree!

\medskip
{\bf t$\bf  _0$ }(Figure 2b): In the 1950's,  the first applications of
stellar evolution models   to determine  ages  for globular   clusters
resulted  in  ages somewhat  older  than the  age  of the Earth; these
estimates climbed considerably to about 26 billion  years in the early
1960's as illustrated  in Figure 2b.  Much  of  this increase resulted
from a change  in  the adopted  helium  abundance (see the  historical
review by Demarque {\it et al.}  1991).  As described by Demarque {\it
et al.}    (and references therein),   the value   of 5 billion  years
obtained by  Sandage in 1953    assumed a helium abundance of   Y=0.4,
whereas the value of 26 billion years obtained  by Sandage in 1962 was
based   on an  adopted  value of  Y=0.1.  The  age estimates began  to
stabilize once  it was   recognized  that  the  helium   abundance for
globular clusters was closer to  that of the  Sun (Y$\sim$0.25).   The
ages of globular clusters have remained at approximately 15-16 billion
years (bracketed loosely   by the bounds  shown  schematically in  the
figure) for some  time; only recently,  with the new results from  the
Hipparcos satellite  (plus new opacities)  have the ages again dropped
systematically. These new results are discussed further below.

These  cartoons illustrate  a  couple  of simple  and  obvious points.
Ultimately, values of  cosmological parameters will not be  determined
by market value  or popular enthusiasm;  they must  come from accurate
experiments.  But as also  indicated in these plots, such measurements
are difficult, they are  dominated by systematic uncertainties, and so
require   a very high  level of  testing, independent measurements and
scrutiny, before  we can know with confidence  if convergence (if any)
is real.

\medskip

\large
\begin{center}
{\bf Determination of $\bf \Omega_m$}
\end{center}
\normalsize
\medskip

On the scale  sizes of clusters  of galaxies {$\sim$ 1  h$^{-1}$ Mpc,
many techniques have been applied to  estimate $\Omega_m$ ({\it e.g.},
Bahcall, this   conference; Bahcall \&  Fan  1998; Dekel,  Burstein \&
White 1997).  The  bottom line  is  that the apparent  matter  density
appears  to  amount  to  only $\sim$20-30\%  of  the  critical density
required for a flat, $\Omega$ = 1  universe.  In fact, the most recent
data are consistent with   most of the   extant data from the  past 20
years.  Cluster velocity dispersions (Carlberg {\it et al.} 1996), the
distortion of   background  galaxies behind  clusters  or weak lensing
(Kaiser \& Squires 1993; Smail {\it et al.}  1995), the baryon density
in clusters (White {\it et al.} 1993), and the  existence of very
massive  clusters at high redshift   (Bahcall  \& Fan), all  currently
favor a low value of the  matter density ($\Omega_m \sim$ 0.2-0.3), at
least on scales up to about  2h$^{-1}$ Mpc.

However,   measurements at scales   larger  than that of clusters  are
extremely challenging,  and determinations of  $\Omega_m$ have not yet
converged  ({\it e.g.},   see  Dekel, Burstein  \&  White  1997).  For
example,  measurements of peculiar velocities   of galaxies, have  led
independent   groups  to  come  to  very  different  conclusions, with
estimates of   $\rm \Omega_m$ ranging from  about  0.2 to 1.3.  Dekel,
Burstein \&  White conclude that the  peculiar  velocity results yield
$\Omega_m >$ 0.3 at the 2-$\sigma$ level.  A new weak lensing study of
a supercluster  (Kaiser {\it et  al.} 1998) on  a  scale of 6 h$^{-1}$
Mpc, yields  a (surprisingly) low  value of  $\Omega_m$ ($\sim$ 0.05),
under the assumption that there is no bias in the way that mass traces
light.    Small  \& Sargent  (1998) have   recently probed  the matter
density for the Corona Borealis supercluster (at a  scale of $\sim$ 20
h$^{-1}$ Mpc), finding $\Omega_m \sim$0.4.  Under  the assumption of a
flat universe, global  limits can also  be  placed on  $\Omega_m$ from
studies of   type  Ia supernovae   (see  next section); currently  the
supernova results favor a value $\Omega_m \sim$ 0.3.

The measurement of the total matter density of the Universe remains an
important and challenging problem. It should be emphasized that all of
the  methods for measuring   $\Omega_m$  are based    on a number   of
underlying assumptions.   For different   methods, the  list  includes
diverse assumptions  about  how  the   mass distribution  traces   the
observed  light distribution,  whether  clusters are representative of
the Universe,  the   properties and  effects of dust    grains, or the
evolution of  the objects  under study.  The   accuracy of any  matter
density estimate  must ultimately be evaluated in   the context of the
validity of the underlying assumptions upon which the method is based.
Hence, it is non-trivial to assign  a quantitative uncertainty in many
cases  but, in fact, systematic effects  (choices and assumptions) may
be the dominant source of uncertainty.

An exciting  result has emerged   this year from  atmospheric neutrino
experiments undertaken  at  Superkamiokande  (Totsuka,  this  volume),
providing evidence for vacuum  oscillations  between muon and  another
neutrino species, and  a  lower limit to  the mass  in neutrinos.  The
contribution of neutrinos to the total density  is likely to be small,
although interestingly it may be comparable to that in stars.

Determining whether   there is     a significant,   smooth  underlying
component to  the matter density on the  largest  scales is a critical
issue that must  be definitively resolved.   If, for example,  some or
all  of  the  non-baryonic dark   matter is   composed of very  weakly
interacting  particles, that component   could prove very  elusive and
difficult   to  detect.  It  unfortunately remains  the   case that at
present, it is not yet  possible to distinguish {\it unambiguously and
definitively} among $\Omega_m$ = 1, $\Omega_m$ + $\Omega_\Lambda$ = 1,
and open  universes with  $\Omega_0  <$ 1,  models all  implying  very
different  underlying fundamental   physics.   The  preponderance   of
evidence at the  present  time, however, does  not  favor the simplest
case of $\Omega_m$ = 1 (the Einstein-de Sitter universe).

\medskip

\large
\begin{center}
{\bf Determination of $\bf \Omega_\Lambda$ }
\end{center}
\normalsize
\medskip

As  illustrated in Figure 2a, the  cosmological constant $\Lambda$ has
had a long and  volatile history in  cosmology.  There have been  many
reasons  to be  skeptical about a  non-zero  value of the cosmological
constant.  To begin with,  there is a  discrepancy of $\geq$120 orders
of magnitude between current observational limits and estimates of the
vacuum energy density based on current  standard particle theory ({\it
e.g.}  Carroll, Press and  Turner 1992).  A  further difficulty with a
non-zero value  for $\Lambda$ is that  it appears coincidental that we
are now living  at a special  epoch when the cosmological constant has
begun to affect the dynamics of the Universe (other than during a time
of  inflation).    It   is also  difficult  to  ignore   the fact that
historically a non-zero  $\Lambda$ has been  called upon to  explain a
number of other apparent crises,  and moreover, adding additional free
parameters to a problem always makes it easier to fit data.

However, despite the strong arguments  have been made for $\Lambda$  =
0, there  are growing  reasons for a  renewed  interest in a  non-zero
value.  Although the  current value of  $\Lambda$ is small compared to
the observed limits, there is no known physical principle that demands
$\Lambda$ =  0 ({\it  e.g.}, Carroll,  Press \&  Turner 1992). Although
Einstein originally introduced   an arbitrary constant  term, standard
particle theory and inflation now provide a physical interpretation of
$\Lambda$:  it  is  the  energy density   of the  vacuum  ({\it e.g.},
Weinberg  1989).  Finally, a number  of  observational results can  be
explained with  a low $\Omega_m$ and  $\Omega_m + \Omega_\Lambda = 1$:
for  instance,   the observed large  scale   distribution of galaxies,
clusters,  and voids described  previously,  in addition to the recent
results  from type  Ia supernovae described   below.  In addition, the
discrepancy between the ages of the oldest stars and the expansion age
(exacerbated if $\Omega_m$ = 1) can be resolved.

Excitement has recently been generated by the  results from two groups
studying type Ia supernovae at high  redshift (one team's results were
reported at  this meeting by  Ariel Goobar).   Both  groups have found
that the high redshift supernovae are fainter (and therefore further),
on  average, than implied  by either an  open  ($\Omega_m$ = 0.2) or a
flat, matter-dominated   ($\Omega_m$ =  1)   universe.  The   observed
differences are $\sim$0.25 and 0.15 mag, (Reiss {\it et al.}  1998 and
Perlmutter  {\it  et  al.}    1998a,  respectively),  or  equivalently
$\sim$13\% and  8\% in distance.  A number  of tests have been applied
to   search for possible systematic    errors that might produce  this
observed effect, but none has  been identified.  Taken at face  value,
these  results imply that  the vacuum energy  density of the Universe,
($\Lambda$), is non-zero.

The early results from these two groups have evolved as more data have
become  available.   Perlmutter {\it et  al.}   (1997)  first reported
results based on a  sample of 7 high-redshift (z$\sim$0.4) supernovae.
Initially, they   found evidence for  a high  matter density $\Omega_m
\sim$ 0.9 $\pm $0.3, with a  value of $\Omega_\Lambda$ consistent with
zero.  However,    with  the  subsequent discovery  of    a z$\sim$0.8
supernova, Perlmutter   {\it et al.}    (1998a) found instead   that a
low-mass density ($\Omega_m  \sim$ 0.2) universe  was  preferred.  The
second, independent group   obtained  preliminary results  based  on 4
supernovae  which  were also consistent   with  a lower matter density
(Garnavich  {\it  et al.}   1998).  The  sample  sizes have  now grown
larger, with 10 supernovae being reported by  Reiss et al.  (1998) and
42 supernovae  being  reported by Perlmutter {\it   et  al.}  (1999).
These two new larger   data sets are yielding consistent  conclusions,
and  the supernovae are now indicating  a  non-zero and positive value
for  $\Omega_\Lambda \sim$ 0.7,  and a small matter density, $\Omega_m
\sim$0.3,  under the  assumption that $\Omega_m$  + $\Omega_\Lambda$ =
1. If a flat universe is  not assumed, the best  fit to the Perlmutter
{\it et al.}   data  yields  $\Omega_m$  =  0.73,  $\Omega_\Lambda$  =
1.32.   The Hubble diagram  for  both the  nearby  (Hamuy {\it et al.}
1996) and the distant (Reiss {\it et al.}  1998) samples of supernovae
are shown in Figure 3.

The advantages of   using  type  Ia  supernovae for    measurements of
$\Omega_\Lambda$   are many.  The dispersion  in   the nearby type  Ia
supernova Hubble diagram is  very small (0.12 mag  or 6\% in distance,
as reported by  Reiss  {\it  et  al.}  1996).   They  are  bright  and
therefore   can  be observed  to   large distances.  In  principle, at
z$\sim$1,  the   shape of the Hubble   diagram  alone can  be  used to
separate $\Omega_m$  and $\Omega_\Lambda$, independent  of the nearby,
local  calibration sample    (Goobar \& Perlmutter  1995).   Potential
effects  due  to evolution, chemical  composition dependence, changing
dust properties are all amenable to empirical tests and calibration.

A possible  weakness of all of  the current supernova $\Omega_\Lambda$
studies  is that the  luminosities of the high-redshift supernovae are
all measured  relative to the same  set of local supernovae.  Although
in the future, estimates of $\Omega_\Lambda$  at high redshift will be
possible using   the shape  of the  Hubble   diagram alone (Goobar  \&
Perlmutter 1995), at present, the  evidence for $\Omega_\Lambda$ comes
from a differential comparison of the nearby sample of supernovae at z
$<$ 1,  with   those  at  z  $\sim$   0.3-0.8.   Hence, the   absolute
calibrations,  completeness levels, and  any other  systematic effects
pertaining  to both datasets are critical.    For several reasons, the
search  techniques   and calibrations  of the   nearby and the distant
samples are  different.  Moreover, the intense  efforts to  search for
high-redshift objects have now led  to the situation where the  nearby
sample is now  smaller than the distant  samples.  While the different
search    strategies  may    not  necessarily  introduce    systematic
differences, increasing  the  nearby sample will provide  an important
check. Such searches are now underway by several groups.

Although a 0.25 mag difference between the  nearby and distant samples
appears large,  the history   of  measurements  of H$_0$   provides an
interesting context   for  comparison.   In     the  case  of    H$_0$
determinations,   a  difference  of  0.25 mag     in zero  point  only
corresponds to  a  difference between 60 and   67 km/sec/Mpc!  Current
differences in the published values for  H$_0$ result from a number of
arcane  factors: the adoption   of different calibrator  galaxies, the
adoption of different techniques for measuring distances, treatment of
reddening and metallicity, and differences in adopted photometric zero
point.    In    fact,  despite  the   considerable     progress on the
extragalactic distance scale  and the  Hubble  constant, recent  H$_0$
values tend  to  range from  about    60 to  80  km/sec/Mpc (see  next
section).   (As recently as  five years ago, there  was  a factor of 2
discrepancy in  these  values, corresponding to   a difference of  1.5
mag.)

In   interpreting the observed   difference between nearby and distant
supernovae, it is also important to  keep in mind  that, for the known
properties  of    dust in  the interstellar   medium,    the  ratio of
total-to-selective absorption, (R$_B$ = A$_B$ / E(B-V)), (the value by
which  the colors are multiplied to  correct the  blue magnitudes), is
$\sim$ 4.  Hence, very accurate photometry and  colors are required to
ultimately understand this issue.   A relative error  of only 0.03 mag
in  color could  contribute  0.12  mag  to the observed  difference in
magnitude.

Further tests and limits on $\Lambda$ may come from gravitational lens
number density statistics (Fukugita  {\it et al.}  1990; Fukugita  and
Turner 1991; Kochanek 1996), plus more stringent limits to the numbers
of close-separation   lenses.   The  numbers  of  strong gravitational
lenses detected depends on the volume surveyed; hence, the probability
that  a quasar   will  be lensed  is a   very   sensitive function  of
$\Omega_\Lambda$.  In  a   flat universe  with $\Omega_\Lambda$  =  0,
almost  an order of  magnitude fewer lenses are   predicted than for a
universe with $\Omega_\Lambda$ =  1.  

If  the current results from  supernovae are correct, then the numbers
of   close-separation   lenses should   be  significantly  larger than
predicted for $\Lambda$ = 0  models. Complications for the lens number
density statistics arise due to a number of  factors which are hard to
quantify  in an  error estimate, and   which become increasingly  more
important  for smaller   values  of $\Lambda$:   for example, galaxies
evolve (and  perhaps  merge)  with time, galaxies  contain   dust, the
properties of the lensing galaxies  are not well-known (in particular,
the dark matter velocity dispersion   is unknown), and the numbers  of
lensing systems for which  this type of  analysis has been carried out
is  still very small.   However, the sample of  known  lens systems is
steadily growing, and new limits from this method will be forthcoming.

The gravitational lens number density  limits from Kochanek (1996) are
$\Omega_\Lambda   <$ 0.66    (95\%   confidence)   for $\Omega_m$    +
$\Omega_\Lambda$ = 1. However, more  recently, Cheng \& Krauss  (1998)
have reinvestigated the sensitivity of this method to various factors.
As Kochanek and Cheng \& Krauss have underscored, the uncertainties in
modelling of  the lensing  galaxies  (generally as  isothermal spheres
with core radii), the observed luminosity functions, core radii of the
galaxies, and the resulting magnification  bias  (that results due  to
the fact  that the lensed   quasar images  are  amplified, and  hence,
easier to detect than if there were no lensing) all need to be treated
carefully.  Also, as Cheng \&  Krauss emphasize, the optical depth for
lensing depends  on the velocity dispersion  to  the fourth  power and
hence, better accuracies  in the  velocity  dispersions are  required.
Cheng \&  Krauss conclude    that systematic uncertainties   currently
dominate the results from this method, but that a flat universe with a
low value of $\Omega_m$ cannot be excluded.

\medskip

\large
\begin{center}
{\bf Determination of $\bf H_0$}
\end{center}
\normalsize
\medskip

The range   of  both previous and current    published values  for the
expansion rate,  or Hubble constant, H$_0$  (see  Figure 1a), attest to
the difficulty of measuring this  parameter accurately.   Fortunately,
the  past  15  years has seen    a series of substantive  improvements
leading toward the  measurement of  a  more accurate value  of  H$_0$.
Indeed, it is quite likely that the 1-$\sigma$ uncertainty in H$_0$ is
now  approaching 10\%, a  significant  advance over the  factor-of-two
uncertainty   that  lingered  for  decades.   Briefly, the significant
progress  can be mainly  attributed to the replacement of photographic
cameras  (used  in  this  context from the   1920's to  the 1980's) by
solid-state detectors, as well as  to both the development of  several
completely new, and the  refinement of existing, methods for measuring
extragalactic  distances  and H$_0$  ({\it   e.g.,} Livio,  Donahue \&
Panagia 1997; Freedman 1997b).

Currently there are many   empirical  routes to the  determination  of
H$_0$;  these fall into the following  completely independent and very
broad categories: 1) the gravitational lens time  delay method, 2) the
Sunyaev-Zel'dovich method   for  clusters, and 3) the    extragalactic
distance scale.  In the latter category, there are several independent
methods  for measuring distances    on the largest   scales (including
supernovae),   but  most of  these    methods share common,  empirical
calibrations at   their  base.  In   the  future,  another independent
determination of H$_0$, from measurements of anisotropies in the cosmic
microwave background,  may also be feasible, if  the physical basis for
the anisotropies can be well-established.

Each of the above methods carries its own susceptibility to systematic
errors,  but the methods as  listed  here, have completely independent
systematics. If history  in this field has  taught us nothing else, it
offers the following important  message:  {\it systematic errors  have
dominated, and continue to dominate, the measurement of H$_0$}.  It is
therefore vital to  measure H$_0$ using  a variety of  methods, and to
test   for the systematics that   are affecting each  of the different
kinds of techniques.

Not  all of these  methods have  yet been tested   to the same degree.
Important progress is being made on  all fronts; however, some methods
are still limited by  sample  size and  small-number statistics.   For
example, method 1), the gravitational  time delay method, has only two
well-studied lens systems  to date: 0957+561 and  PG 1115.  The  great
advantage of both  methods 1) and  2),  however, is that  they measure
H$_0$ at very large distances,  independent of the  need for any local
calibration.

\medskip
\subsection{ 1) Gravitational Lenses}
\medskip

Refsdael (1964, 1966) noted that the  arrival times for the light from
two gravitationally lensed images   of a background point  source  are
dependent   on the  path  lengths  and   the  gravitational  potential
traversed in each case. Hence, a measurement of the time delay and the
angular separation for different images   of a variable quasar can  be
used to  provide a  measurement  of  $\rm H_0$.   This  method  offers
tremendous potential because it can be applied  at great distances and
it is   based  on  very   solid physical principles    (Blandford  \&
Kundi\'{c} 1997).

There are  of course difficulties with  this method  as there are with
any other.    Astronomical   lenses  are galaxies    whose  underlying
(luminous or dark) mass distributions are not independently known, and
furthermore they may  be sitting in more  complicated group or cluster
potentials.  A degeneracy exists between the  mass distribution of the
lens and the  value of H$_0$ ({\it  e.g.,}  Keeton and Kochanek  1997;
Schechter  {\it    et  al.}   1997).   Ideally     velocity dispersion
measurements as a  function of position are  needed (to  constrain the
mass distribution of the lens).   Such measurements are very difficult
(and generally have   not been  available).   Perhaps  worse yet,  the
distribution of the dark matter in these systems is unknown.

Unfortunately, to  date, there are very few  systems  known which have
both a favorable geometry (for providing  constraints on the lens mass
distribution) and a  variable background source (so  that a time delay
can be measured). The two systems to date  that have been well-studied
yield  values of H$_0$  in the approximate   range of 40-70 km/sec/Mpc
(Schechter {\it et  al.}   1997; Impey {\it  et  al.}  1998) with   an
uncertainty of $\sim$20-30\%.  These values assume a value of $\Omega$
= 1, and  rise by 10\%  for low $\Omega$.  Tonry  \& Franx (1998) have
recently reported an accurate   new velocity dispersion of  $\sigma$ =
288 $\pm$ 9 km/sec  for the lensing galaxy  in 0957+561, based on data
obtained  at the Keck 10m  telescope.  Adopting  $\Omega_m$ = 0.25 and
the model of Grogin \& Narayan (1996) for the mass distribution of the
lens yields a  value of  H$_0$ =  72 $\pm$  7 (1-$\sigma$ statistical)
$\pm$ 15\% (systematic).

As the number of  favorable lens systems  increases (as further lenses
are discovered  that have measurable time  delays), the  prospects for
measuring  H$_0$ and   its    uncertainty using  this    technique are
excellent.  Schechter (private communication)  reports that there  are
now 6 lenses with measured time delays, but perhaps only half of these
will  be useful  for H$_0$   determinations due to  the difficulty  of
modelling the galaxies.

\medskip
\subsection{ Sunyaev Zel'dovich Effect and X-Ray Measurements}
\medskip

The inverse-Compton scattering   of photons from the cosmic  microwave
background   off of hot  electrons in  the  X-ray gas of rich clusters
results in a measurable decrement in the microwave background spectrum
known  as the Sunyaev-Zel'dovich  (SZ)  effect (Sunyaev \&  Zel'dovich
1969).    Given a   spatial distribution    of the   SZ   effect and a
high-resolution X-ray map,  the density  and temperature distributions
of the hot  gas can be obtained;  the mean electron temperature can be
obtained from an X-ray spectrum.   The method  makes  use of the  fact
that     the  X-ray flux       is  distance-dependent,    whereas  the
Sunyaev-Zel'dovich decrement in the temperature is not.

Once again, the advantages  of this method  are that it can be applied
at  large  distances   and, in principle,  it    has a straightforward
physical basis.  Some of the main  uncertainties result from potential
clumpiness  of  the gas (which   would result in  reducing $\rm H_0$),
projection effects (if  the clusters observed  are prolate,  $\rm H_0$
could  be larger), the  assumption of hydrostatic equilibrium, details
of the models    for the gas   and electron  densities, and  potential
contamination from point sources.

To date, a range of values  of $\rm H_0$  have been published based on
this  method ranging from     $\sim$40 - 80  km/sec/Mpc  ({\it  e.g.},
Birkinshaw 1998).  The uncertainties are  still large, but as more and
more clusters are    observed, higher-resolution  (2D)  maps of    the
decrement, and X-ray maps  and spectra become available, the prospects
for this method are improving enormously.

Carlstrom, (this meeting) presented exquisite  new measurements of the
Sunyaev-Zel'dovich decrement,  measured   in  two  dimensions  with an
interferometer, for a number  of nearby clusters.  He also highlighted
the imminent  advances on the horizon for  X-ray  imaging using NASA's
soon-to-be launched Chandra  X-ray Observatory (the satellite formerly
known  as  AXAF).   There  is  promise of  a   significant increase in
accuracy for this method in the near future.

\medskip
\subsection {The Extragalactic Distance Scale}
\medskip

The launch of the  Hubble Space Telescope  (HST) in 1990  has provided
the  opportunity  to  undertake  a  major  program  to calibrate   the
extragalactic distance scale.   The resolution of  HST is  an order of
magnitude higher than  can be generally  obtained through  the Earth's
atmosphere, and  moreover  it is  stable; as a  result,  the volume of
space made accessible by HST increased by 3 orders  of magnitude.  The
HST Key  Project on the  extragalactic distance scale was  designed to
measure the Hubble  constant  to   an  accuracy of $\pm$10\%     $rms$
(Freedman {\it et al.}  1994; Mould {\it et al.}  1995; Kennicutt {\it
et al.}  1995).  Since the dominant sources of error are systematic in
nature, the approach we have  taken in the Key  Project is to  measure
H$_0$  by   intercomparing several different   methods.  This approach
allows us  to assess  and  quantify explicitly the systematic  errors.
The HST Key Project will be completed  in 1999; since new results will
be available  shortly,  this   discussion  will be kept   very  brief.
Results based on half  of the available data yield  H$_0$ = 72 $\pm$ 5
$\pm$ (random) 7 (systematic) km/sec/Mpc  (Madore {\it et al.}   1998,
1999; Freedman {\it et   al.}  1998; Mould  {\it et  al}.   1997).  In
Figure 4,  the results for various   H$_0$ methods are  combined using
both a Frequentist and a Bayesian approach (from Madore {\it et al.}).

The largest  remaining  sources  of uncertainty  in the  extragalactic
distance route to H$_0$  can be traced to  uncertainty in the distance
to the Large Magellanic Cloud (the galaxy  which provides the fiducial
comparison for more distant galaxies), and to the potential effects of
differing amounts  of elements heavier  than helium  (or metallicity).
The  importance of the latter  effect has been difficult to establish.
The  recently-installed infrared (NICMOS) camera  on HST is being used
to address this, and may help to resolve the issue shortly.

A histogram  of the  distribution of  distances to  the LMC   from the
literature is shown in Figure 5.   The distances in this histogram are
based on Cepheids, RR Lyraes, SN 1987A, red giants, the ``red clump'',
and long-period  variables.    Values  prior  to 1996  come   from the
published   compilation of   Westerlund (1997),  but  only  the latest
revision published by a given author is plotted for  a given data set.
Despite decades  of effort  in measuring  the distance  to this nearby
neighboring galaxy, and the  number of independent  methods available,
the   dispersion   in    measured  distance   modulus   remains   very
high. Moreover, the distribution is not Gaussian.  There has been much
recent activity on the red clump which  contributes many of the values
around  18.3 mag,  and  gives rise   to  the  bimodal  nature  of  the
distribution.  There  is as yet  no understanding  of  why there is  a
systematic difference  between the Cepheid  and the red clump distance
scale.    This  histogram illustrates   that  the  uncertainty  in the
distance  to the  LMC  is  still  large.   Without assuming that   the
distribution is Gaussian, the 95\% confidence limits are +/- 0.28 mag,
and  the 68\%   confidence limits amount   to +/-  0.13  mag  or 7
distance. Unfortunately, the distance to the LMC remains as one of the
largest systematic uncertainties in the current extragalactic distance
scale.

\medskip

\large
\begin{center}
{\bf Determination of t$_0$}
\end{center}
\normalsize
\medskip

Age-dating of  the  oldest  known objects  in the   Universe  has been
carried out in a number of ways.  The most reliable ages are generally
believed to come from the application of theoretical models of stellar
evolution to observations of the oldest clusters in the Milky Way, the
globular  clusters. For about 30 years,  the ages of globular clusters
have remained reasonably   stable, at about   15  billion years  ({\it
e.g.}, Demarque {\it et  al.} 1990; Vandenberg  {\it  et al.}   1996);
however, recently these ages have  been revised downward, as described
below.    Ages  can  also  be  obtained  from    radioactive dating or
nucleocosmochronology ({\it e.g.}, Schramm  1989), and a further lower
limit can  be estimated from the  cooling rates for white dwarfs ({\it
e.g.}, Oswalt  {\it  et al.} 1996).    Generally, these ages have
ranged from about 10 to 20 billion years; the largest sources of
uncertainty in each of these estimates are  again systematic in nature.

During the 1980's and 1990's, the  globular cluster age estimates have
improved as both new observations of globular  clusters have been made
with CCD's, and as refinements  to stellar evolution models, including
updated  opacities, consideration  of  mixing, and  different chemical
abundances, have   been incorporated ({\it  e.g.}, Vandenberg  {\it et
al.}  1996; Chaboyer {\it et al.}   1996, 1998). The latter authors
have undertaken  an  extensive analysis of  the uncertainties   in the
determination  of  globular  cluster  ages.   From  the  theory  side,
uncertainties  in globular   cluster ages result,   for example,  from
uncertainties in how convection is treated, the opacities, and nuclear
reaction rates.  From the measurement  side uncertainties arise due to
corrections for    dust and chemical    composition; however, {\it the
dominant source of overall uncertainty in the globular cluster ages is
the uncertainty  in the cluster   distances}.  

In  fact, the impact  of distance uncertainty  on the ages of globular
clusters   is twice as large  as  its effect  on  the determination of
H$_0$.  That is, a 0.2  mag difference in  zero point corresponds to a
10\% difference in the  distance (or  correspondingly, H$_0$), but  it
results  in a  20\% difference in  the age  of  a cluster  ({\it e.g.},
Renzini 1991).

The   Hipparcos satellite has   recently  provided geometric  parallax
measurements for 118,000 nearby stars (Kovalevsky 1998).  Relevant for
the  calibration  of globular cluster    distances, are the relatively
nearby old  stars of  low  metal composition,  the  so-called subdwarf
stars,  presumably the nearby analogs of  the old, metal-poor stars in
globular  clusters.    Accurate distances to  these   stars  provide a
fiducial calibration   from  which   the    absolute luminosities   of
equivalent stars in  globular clusters can  be determined and compared
with  those   from  stellar  evolution  models.    The   new Hipparcos
calibration has  led to  a  downward revision of the  globular cluster
ages from  $\sim$15 billion years to 11-14  billion years  ({\it e.g.},
Reid 1997; Pont {\it et al.  } 1998; Chaboyer {\it et al.}  1998).

However, as  emphasized by Chaboyer  {\it  et al.},  there are only  8
stars in the Hipparcos catalog having both  small parallax errors, and
low metal abundance,  [Fe/H] $<$-1.1 (i.e.,   less than one  tenth the
iron-to-hydrogen abundance relative  to  the solar value).   In  fact,
Gratton {\it  et  al.}   (1998) note  that  there  are no   stars with
parallax errors $<$10\%  with [Fe/H]$<$-2 corresponding to the oldest,
metal  poor globular clusters.   Hence,  the  calibration of  globular
cluster ages based  on parallax measurements  of old, metal-poor stars
remains an area where an increase  in sample size  will be critical to
beat down the statistical uncertainties.  A decade or so from now, new
parallax  satellites such as   NASA's  SIM (the  Space  Interferometry
Mission) and the European Space Agency's mission  (named GAIA) will be
instrumental in improving these  calibrations, not only for subdwarfs,
but for many  other  classes of stars (for  example,  Cepheids and the
lower-mass variable  RR Lyrae stars).   These interferometers  will be
capable of delivering  2--3    orders  of magnitude   more    accurate
parallaxes than   Hipparcos,  down  to  fainter  magnitude  limits for
several orders  of  magnitude  more  stars.  Until larger  samples  of
accurate parallaxes are available, however, distance errors are likely
to continue to contribute the largest source of systematic uncertainty
to the globular cluster ages.


\medskip
\large
\begin{center}
{\bf The Cosmic Microwave Background Radiation and Cosmological Parameters }
\end{center}
\normalsize
\medskip

As discussed at this meeting by  Silk, Wilkinson and Spergel, over the
next few  years, increasingly more accurate  measurements will be made
of   the  fluctuations  in the    cosmic   microwave background  (CMB)
radiation.  The  underlying physics  governing the  shape of   the CMB
anisotropy  spectrum can be  described  by the  interaction of a  very
tightly coupled fluid   composed of   electrons and photons     before
recombination ({\it   e.g.}, Hu \& White   1996; Sunyaev \& Zel'dovich
(1970). Figure 6 shows a plot  of the predicted angular power spectrum
for CMB anisotropies from Hu,  Sugiyama \& Silk (1997), computed under
the assumption that the fluctuations  are Gaussian and adiabatic.  The
position  of the first  angular peak is  very  sensitive to $\Omega_0$
($\Omega_m$ + $\Omega_\Lambda$ + $\Omega_k$).

For information on  cosmological  parameters to be extracted  from the
CMB  anisotropies, the  following must  be  true: first,  the physical
source of these fluctuations  must   be understood, and second,    the
sources of  systematic error must be eliminated  or minimized  so that
they do not dominate the uncertainties.  

Recently  it  has become clear that  almost  exact  degeneracies exist
between various  cosmological  parameters ({\it e.g.},   Efstathiou \&
Bond 1998;  Eisenstein, Hu \&  Tegmark  1998) such that,  for example,
cosmological models with the same matter density can have the same CMB
anisotropies,  while having very   different geometries. As a  result,
measurement of  CMB anisotropies will, in  principle, be able to yield
strong   constraints   on     the    products   $\Omega_m$h$^2$    and
$\Omega_b$h$^2$, but not  on the individual values  of h (= H$_0$/100)
and $\Omega_m$   directly.   Hence,  earlier    suggestions that  such
cosmological  parameters  could be measured  from  CMB anisotropies to
precisions of  1\% or better  ({\it e.g.}, Bond, Efstathiou \& Tegmark
1997) will unfortunately  not  be realized.  However,   breaking these
degeneracies can be accomplished by  using the CMB data in combination
with other data, for example, the Sloan survey and type Ia supernovae
({\it e.g.} White 1998).

Currently  the estimates  of  the  precisions for which   cosmological
parameters can  be extracted from measurements  of anisotropies in the
CMB  are  based on  models  in  which the  primordial fluctuations are
Gaussian  and adiabatic,  and for which   there is no preferred scale.
Very detailed predictions can be made for this model, more so than for
competing models  such as  isocurvature  baryons or cosmic strings  or
textures.   In the next few  years, as the  data improve, all of these
models will be scrutinized in greater detail.

Important    additional  constraints   may    eventually   come  from
polarization measurements  ({\it e.g.}, Zaldarriaga, Spergel \& Seljak
1997; Kamionkowski {\it et al.}  1997), but these may require the next
generation of experiments beyond MAP and Planck. The polarization data
may provide a means of breaking some  of the degeneracies amongst the
cosmological  parameters   that are  present in   the temperature data
alone. Furthermore,  they are sensitive  to the  presence of a  tensor
(gravity wave) contribution, and hence can allow a very sensitive test
of competing models.

Although it  is  not  yet  certain  how  accurately  the  cosmological
parameters can be extracted   from measurements of  CMB  anisotropies,
what is   clear  is  that   upcoming, scheduled   balloon   and  space
experiments   offer an opportunity  to  probe detailed  physics of the
early  Universe.   If current models  are  correct, the first acoustic
peak   will be confirmed very  shortly    and its position  accurately
measured by balloon experiments even  before the launch of MAP.  These
balloon experiments  will soon  be  followed  with the total   sky and
multi-frequency coverage provided by MAP and Planck.  This new era now
being  entered, of precision  CMB anisotropy experiments, is extremely
exciting.

\medskip

\large
\begin{center}
{\bf Discussion and Summary }
\end{center}
\normalsize
\medskip

In the past year, a radical shift has begun to occur.  Until recently,
a  majority of the      theoretical community viewed    the (standard)
Einstein-  de Sitter model ($\Omega_0$ =   1, $\Omega_\Lambda$ = 0) as
the most likely  case (with h=  0.5, t  = 13  Gyr).  With accumulating
evidence for  a low matter  density, difficulty in fitting  the galaxy
power spectrum with  the standard model, the conflict  in ages for the
Einstein-de Sitter  case, and now,   most recently, the evidence  from
type Ia supernovae for   an accelerating universe, a new  ``standard''
model  is emerging, a model with  $\Omega_m \sim $ 0.3, $\Omega_\Lambda
\sim $ 0.7,  h = 0.65,  and t = 13  Gyr.  This model preserves  a flat
universe and is still consistent with inflation.

In Figure  7,  the  bounds  on  several   cosmological parameters  are
summarized in a plot of the matter density as a function of the Hubble
constant.   What do these   current limits on cosmological  parameters
imply about the contribution of non-baryonic dark  matter to the total
matter   density?  As can  be  seen from  the figure,  for  H$_0$ = 70
km/sec/Mpc, current limits  to   the deuterium abundance  (Burles \&
Tytler  1998;  Hogan  1998) yield  baryon densities  in   the range of
$\Omega_b$ =  0.02 to 0.04,  or 2-4\% of  the critical density.  Given
current  estimates   of the   matter  density ($\Omega_m   \sim $0.3),
non-baryonic matter would thus contribute  just over 25\% of the total
energy density needed for a flat, $\Omega$ = 1 universe.

One might ask, is non-baryonic dark matter still required if $\Lambda$
is non-zero?   Allowance for $\Lambda$  $\neq$ 0 does  not provide the
missing energy to simultaneously yield $\Omega$  = 1, while doing away
with  the necessity of  non-baryonic dark   matter,  at least for  the
current limits  from Big  Bang nucleosynthesis.   As can  be seen from
Figure 7, for  the current deuterium limits,  having all baryonic mass
plus $\Lambda$ would  require both H$_0$ $\sim$  30 km/sec/Mpc and  an
age for the Universe of $\sim$ 30 Gyr.  These H$_0$ and age values are
outside  the  range of currently  measured   values for  both of these
parameters.  Although it might be appealing  to do away simultaneously
with one type of  unknown (non-baryonic dark matter) while introducing
another parameter ($\Lambda$), a  non-zero value for the  cosmological
parameter  does   not remove  the  requirement   for non-baryonic dark
matter.

The question of the nature of dark matter (or energy) remains with us.
In this sense, the situation has  not changed very  much over the past
few decades, although the  motivation for requiring a critical-density
universe  has  evolved from  considerations  of fine-tuning and causal
arguments  to the  development  of inflation.  But  searches for  dark
matter since the 1970's have not uncovered sufficient matter to result
in a critical-density universe.    This year has offered  exciting new
(and therefore  still tentative) results that  a non-zero value of the
cosmological constant,  or    perhaps an evolving  scalar   field like
quintessence (Steinhardt,  this  volume; Steinhardt \& Caldwell  1998)
could provide the  missing  energy  to  result in a   critical-density
universe.   Still,  the   nature  of the   dark    matter,  whether it
contributes 25\% or 95\% of the  total energy density, is unknown, and
remains as one of the most fundamental unsolved problems in cosmology.

The progress in measuring cosmological parameters has been impressive;
still, however, the  accurate measurement  of cosmological  parameters
remains a challenging task.   It is therefore  encouraging to note the
wealth of new data that will appear  over the next few years, covering
very diverse  areas of parameter space.    For example, measurement of
CMB anisotropies, (from balloons and space  with MAP, and Planck), the
Sloan Digital  Sky  Survey, Hubble   Space  Telescope, Chandra   X-ray
Observatory,  radio   interferometry,  gravitational lensing  studies,
weakly  interacting   massive  particle  (WIMP)  cryogenic  detectors,
neutrino experiments, the  Large Hadron Collider (LHC),  and a host of
others, inspire optimism that the noose  on cosmological parameters is
indeed   tightening.  At the   very least, the   next few years should
continue to be interesting!

\section*{Acknowledgments} 
It is a great pleasure to thank the organizers of this Nobel Symposium
for a particularly enjoyable and stimulating conference, and for their
immense hospitality. I also thank Brad Gibson for his help with
the LMC distance literature search.

\def\aj{\em Astron. J.}
\def\araa{\em Astron. Rev. Astron. Astrophys.}
\def\apj{\em Astrophys. J.}
\def\apjl{\em Astrophys. J. Lett.}
\def\apjs{\em Astrophys. J. Suppl.}
\def\apss{\em Astrophys. \& SS}
\def\aa{\em Astron. Astrophys.}
\def\aas{\em Astron. Astrophys. Suppl.}
\def\mnras{\em Mon. Not. Royal Astr. Soc.}
\def\nat{\em Nature,}
\def\pasp{\em Publ. Astr. Soc. Pac.}
\def\physrep{\em Phys. Rep.}
\def\physrev{\em Phys. Rev.}
\def\pnas{\em Publ. Nat. Acad. Sci.}
\def\rmp{\em Rev. Mod. Phys.}
\def\sci{\em Science,}
\def\helv{\em Helv. Phys. Acta}
\def\iau{\em IAU Trans.}


\medskip
\medskip
{\bf References}
\medskip

W. Baade, IAU Trans. {\bf 8}, 397, 1954.

N. A. Bahcall, L. M.  Lubin \& V. Dorman, {\it Astrophys. J. Lett.},  {447, 81, 1995}.

N. A. Bahcall \& Fan, {\it Publ. Nat. Acad. Sci.}, {95, 5956, 1998}.

M. Birkinshaw  {\it Phys. Rep.},  1999, in press.

R. Blandford \& T. Kundi\'{c}, in 
{\em The Extragalactic Distance Scale} eds. M. Donahue \& 

M. Livio (Cambridge University Press, 1997), pp. 60--75.

A. Bosma, {\it Astron. J.},{86, 1791, 1981}.

S. Burles \& D. Tytler, {\it Astrophys. J.}, 507, 732, 1998.

Carlberg, R. G., {\it et al.}, {\it Astrophys. J.}, 462, 32, 1996.

Carroll, Press, \& Turner, {\it Astron. Rev. Astron. Astrophys.},{30, 499, 1992}.

B. Chaboyer, P. Demarque, P. J. Kernan \& L. M. Krauss, {\it Science}, {271, 957, 1996}.

B. Chaboyer, P. Demarque, P. J. Kernan \& L. M. Krauss, {\it Astrophys. J.}, 494, 96, 1998.

Y.-C. N. Cheng \& L. M. Krauss {\it Mon. Not. Royal Astr. Soc.}, 1999}. astro-ph 9810393, in press.

A. Dekel {\it et al.}, {\it Astrophys. J.}, 412, 1, 1993.

A. Dekel, D. Burstein \& S. D. M. White in {\em Critical Dialogs in Cosmology}, ed. N. Turok 

(World Scientific, 1997).

P. Demarque, C. P.  Deliyannis \& A. Sarajedini, in {\em
Observational   Tests  of  Inflation},   

ed.  T. Shanks  {\it  et al.}
(Dordrecht, Kluwer, 1991).

M. Donahue \& M. Livio, eds.  {\em The Extragalactic Distance Scale}

(Cambridge University Press, 1997).

G. Efstathiou \& J. R. Bond, {\it Mon. Not. Royal Astr. Soc.}, 1998,astro-ph/9807103.

A. Einstein, {\em Sitz. Preuss. Akad. d. Wiss. Phys.-Math}, 142, (1917).

D. Eisenstein, W. Hu \& M. Tegmark, {\it Astrophys. J.}, 504, L57, 1998.

W. L. Freedman  in {\em Proceedings of the 18th Texas Symposium},

eds. A. Olinto, J. Frieman, \& D. Schramm, (World Scientific Press, 1997a).

W. L. Freedman  in {\em Critical Dialogs in Cosmology}, 
ed. N. Turok 

(World Scientific, 1997b), p. 92.

W. L. Freedman, B. F. Madore \& R. C. Kennicutt,  in 
{\em The Extragalactic Distance 

Scale} eds. M. Donahue \& M. Livio,
(Cambridge University Press, 1997), pp. 171-185.

W. L. Freedman, J. R. Mould, R. C. Kennicutt \&
B.F. Madore, in IAU Symposium No. 183, 

Cosmological Parameters
and the Evolution of the Universe, in press, 1998, astro-ph/9801080.

A. Friedmann, {\it Z. Phys.}, {\bf 10},  377, (1922).

M. Fukugita, T. Tutamase \& M. Kasai, {\it Mon. Not. Royal Astr. Soc.},{246, 24p, 1990}.

M. Fukugita \& E. Turner, {\it Mon. Not. Royal Astr. Soc.},{253, 99, 1991}.

P. Garnavich {\it et al.}, {\it Astrophys. J.}, 493, L3, 1998.

A. Goobar \& S. Perlmutter, {\it Astrophys. J.}, 450, 14, 1995.

N. A. Grogin \& R. Narayan, {\it Astrophys. J.}, 464, 92, 1996.

A. H. Guth, {\it Phys. Rev.}, {23, 347, 1981}.

C. Hogan, {\it Space Sci. Review}, {\bf 84}, 127 (1998).

W. Hu \& M. White {\it Astrophys. J.}, 441, 30, 1996.

E. Hubble, {\it Publ. Nat. Acad. Sci.}, {15, 168, 1929}.

E. Hubble \& M. L. Humason   {\it Astrophys. J.,} {74, 43, 1931}.

C. Impey, {\it Astrophys. J.}, 509, 551, 1998.

N. Kaiser \& G. Squires, {\it Astrophys. J.}, 404, 441, 1993.

N. Kaiser {\it et al.}, {\it Astrophys. J.,} 1999, astro-ph 9809269, in press.

M. Kamionkowski, A. Kosowsky \& A. Stebbins, astro-ph/9609132.

R. C. Kennicutt, W. L. Freedman \& J. R. Mould, {\it Astron. J.}, {110, 1476, 1995}.

C. S. Kochanek, {\it Astrophys. J.}, 466, 638, 1996.

J. Kovalevsky, {\it Astron. Rev. Astron. Astrophys.}, {36, 99, 1998}.

L. Krauss \& M. S. Turner, {\em Gen. Rel. Grav.}, 27, 1137, 1995.

C.-P. Ma, T. Small \& W. Sargent  1998,
astro-ph/9808034.

B. F. Madore {\it et al.}, {\it Nature}, {395, 47, 1998}.

B. F. Madore {\it et al.}, {\it Astrophys. J.},  1999, in press, Apr. 10 issue, astro-ph/9812157.

J. R. Mould, S. Sakai \& S. Hughes,  in 
{\em The Extragalactic Distance Scale} 

eds. M. Donahue \& M. Livio
(Cambridge University Press, 1997), pp. 158--170.

J. P. Ostriker \&  P. Steinhardt, {\it Nature}, {377, 600, 1995}.

J.    P.  Ostriker, P.    J.  E.  Peebles   \& A.  Yahil
{\it Astrophys. J. Lett.}, 193, L1, 1974.

T. D. Oswalt, J. A. Smith, M. A. Wood \& P. Hintzen, {\it Nature}, {382, 692, 1996}.

S. Perlmutter {\it et al.}, {\it Astrophys. J.}, 483, 565, 1997.

S. Perlmutter {\it et al.}, {\it Nature}, {391, 51, 1998a}.

S. Perlmutter {\it et al.}, {\it Astrophys. J.},  1999,
astro-ph/9812133.

V. Petrosian, E. Salpeter \& P. Szekeres, {\it Astrophys. J.,}
{147, 1222, 1967}.

F. Pont, M. Mayor, C. Turon \& D. A. Vandenberg {it Astron. Astrophys.},
1999, in press.

S. Refsdael, {\it Mon. Not. Royal Astr. Soc.}, {128, 295, 1964}.

S. Refsdael, {\it Mon. Not. Royal Astr. Soc.}, {132, 101, 1966}.

N. Reid, {\it Astron. J.}, 114, 161, 1997.

A. Reiss, W. Press \& R. Kirshner {\it Astrophys. J.}, 473, 88, 1996.

A. Reiss {\it et al.} {\it Astron. J.}, 116, 1009, 1998.

A. Renzini, in {\em Observational Tests of Inflation}, 
ed. T. Shanks {\it et al.} 

(Dordrecht, Kluwer, 1991), pp. 131-146.

D. H. Rogstad \& G. S. Shostak, {\it Astrophys. J.}, 176, 315, 1972.

M. Rowan-Robinson,   {\em The Cosmological Distance Ladder},
(New York, Freeman, 1985).

V. C. Rubin, N. Thonnard \& W. K. Ford, {\it Astrophys. J. Lett.}, 225, L107, 1978.

A. Sandage {\it et al.}, {\it Astrophys. J. Lett.}, {460, 15, 1996}.

A. Sandage, {\it Astrophys. J.}, 127, 513, 1958.

P. Schechter {\it et al.}, {\it Astrophys. J. Lett.}, {475, 85, 1997}.

D. N. Schramm, in {\em Astrophysical Ages and Dating Methods}, 

eds. E. Vangioni-Flam {\it et al.} (Edition Frontieres: Paris, 1989).

I. Smail, R. S. Ellis, M. J. Fitchett \& A. C. Edge, {\it Mon. Not. Royal Astr. Soc.}, {273, 277, 1995}.

P. Steinhardt \& J. Caldwell, in 
{\em The Cosmic Microwave Background and Large Scale Structure 

in the Universe} eds. Y.. I. Byan \& K. W. Ng
(ASP Conf. Series 151, 1998), p. 13.

P. Stetson {\it Publ. Astr. Soc. Pac.}, {110, 1448, 1998}.

R. A. Sunyaev \& Y. B. Zel'dovich, {\it Astrophys. \& SS}, {4, 301, 1969}.

R. A. Sunyaev \& Y. B. Zel'dovich, {\it Astrophys. \& SS}, 7, 3, 1970.

J. Tonry \& M. Franx, astro-ph 9809064.

V. Trimble, {\it Astron. Rev. Astron. Astrophys.}, {25, 425, 1987}.

S. van den Bergh, in 
{\em The Extragalactic Distance Scale} eds. M. Donahue 

\& M. Livio
(Cambridge University Press, 1997), pp. 1--5.

D. A. Vandenberg, M. Bolte, \& P. B. Stetson, {\it Astron. Rev. Astron. Astrophys.}, 34, 461, 1996.

S. Weinberg, {\it Rev. Mod. Phys.}, {61, 1, 1989}.

B. E. Westerlund,  The Magellanic Clouds, (Cambridge:
Cambridge Univ. Press,  (1997).

S. D. M. White, J. F. Navarro, A. E. Evrard \& C. S. Frenk, {\it Nature}, {366, 429, 1993}.

M. White, {\it Astrophys. J.}, 1999.

M. Zaldarriaga, D. N. Spergel \& U. Seljak, {\it Astrophys. J.}, 488, 1, 1997.

F. Zwicky, {\it Helv. Phys. Acta}, {6, 110, 1933}.


\vfill\eject

\begin{center}
{\bf Figure Captions }
\end{center}
\normalsize
\medskip

{\bf Figures  1a-b:} The trend  with time for  measurements  of H$_0$ and
$\Omega_m$. See text for details.

\medskip
{\bf  Figures  2a-b:} The trend  with  time  for $\Lambda$, and t$_0$.
Note the arbitrary units for $\Lambda$. See text for details.

\medskip

{\bf Figure 3 (top panel):} The Hubble  diagram for type Ia supernovae
from Hamuy {\it et al.}  (1996) and Reiss {\it et al.} (1998). Plotted
is the  distance modulus in  magnitudes  versus the logarithm  of  the
redshift. Curves for various cosmological  models are indicated.  {\bf
(bottom panel):} Following Reiss {\it  et al.}  (1998), the difference
in  magnitude  between the observed data   points  compared to an open
($\Omega_m$ = 0.2) model is shown.  The distant supernovae are fainter
by 0.25 magnitudes, on average, than the nearby supernovae.

\medskip




\medskip

{\bf Figure 4}:  Plot of various  H$_0$ determinations and the adopted
values from Madore   {\it et al.}  (1998).   In  the left  panel, each
value  of $H_0$  and its statistical  uncertainty  is represented by a
Gaussian of unit area (linked dotted line)  centered on its determined
value  and having a   dispersion  equal to  the  quoted random  error.
Superposed  immediately above  each  Gaussian   is  a  horizontal  bar
representing the one sigma  limits of the calculated systematic errors
derived  for that determination.    The adopted average value  and its
probability distribution function   (continuous  solid  line)  is  the
arithmetic    sum  of   the individual   Gaussians.   This Frequentist
representation treats  each determination as  independent, and assumes
no  {\it a priori}   reason to prefer  one solution  over another.   A
Bayesian representation of  the  products of the   various probability
density  distributions  is shown in the  right  panel.  Because of the
close proximity and    strong   overlap in the    various  independent
solutions   the  Bayesian estimator  is   very similar to,  while more
sharply defined than, the Frequentist solution.

\medskip

{\bf Figure 5}: A histogram of  distance moduli determinations for the
Large Magellanic  cloud.  Values prior  to 1996  are  from a published
compilation by Westerlund (1997). 

\medskip


{\bf Figure 6}:  The    angular   power spectrum  of    cosmic
microwave      background anisotropies   assuming  adiabatic,   nearly
scale-invariant  models for a  range  of values of $\rm  \Omega_0$ and
$\rm  \Omega_\Lambda$ (Hu, Sugiyama,  and  Silk 1997; their Figure  4).
The C$_l$ values correspond to  the squares of the spherical harmonics
coefficients.   Low $l$ values  correspond to large angular scales ($l
\sim { 200\deg \over  \theta } $).  The position of the first acoustic
peak  is  predicted to be  at  $l  \sim $220$\Omega_{TOT}^{-1/2}$, and
hence, shifts to smaller angular scales for open universes.

\medskip

{\bf Figure  7}:   Plot of $\rm \Omega_m$   versus  $\rm  H_0$ showing
current observational  limits on cosmological  parameters.  The shaded
box is defined by values of H$_0$ in the range of  40 to 90 km/sec/Mpc
and  0.15 $\rm <   \Omega_m  <$ 0.4.   The  thick solid  lines  denote
expansion ages for an open ($\rm \Omega_\Lambda$ = 0) Universe for 10,
15, and 20 Gyr and the thick dashed lines denote expansion ages in the
case of  a flat ($\rm \Omega_m +  \Omega_\Lambda $  =1) Universe.  The
light dashed  lines denote current limits  for $\Omega_b$ based on low
and high values for the deuterium to-hydrogen ratio.



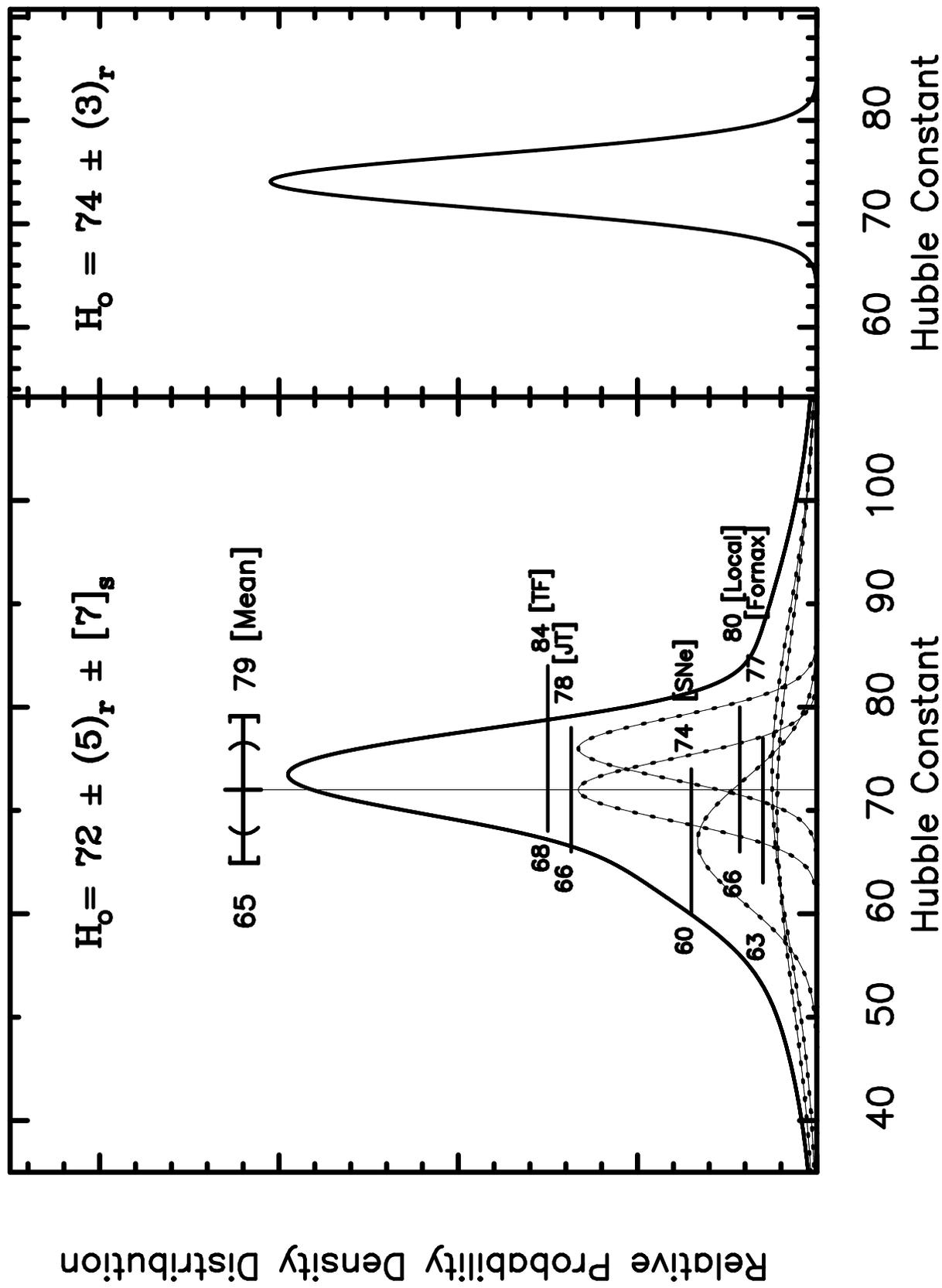
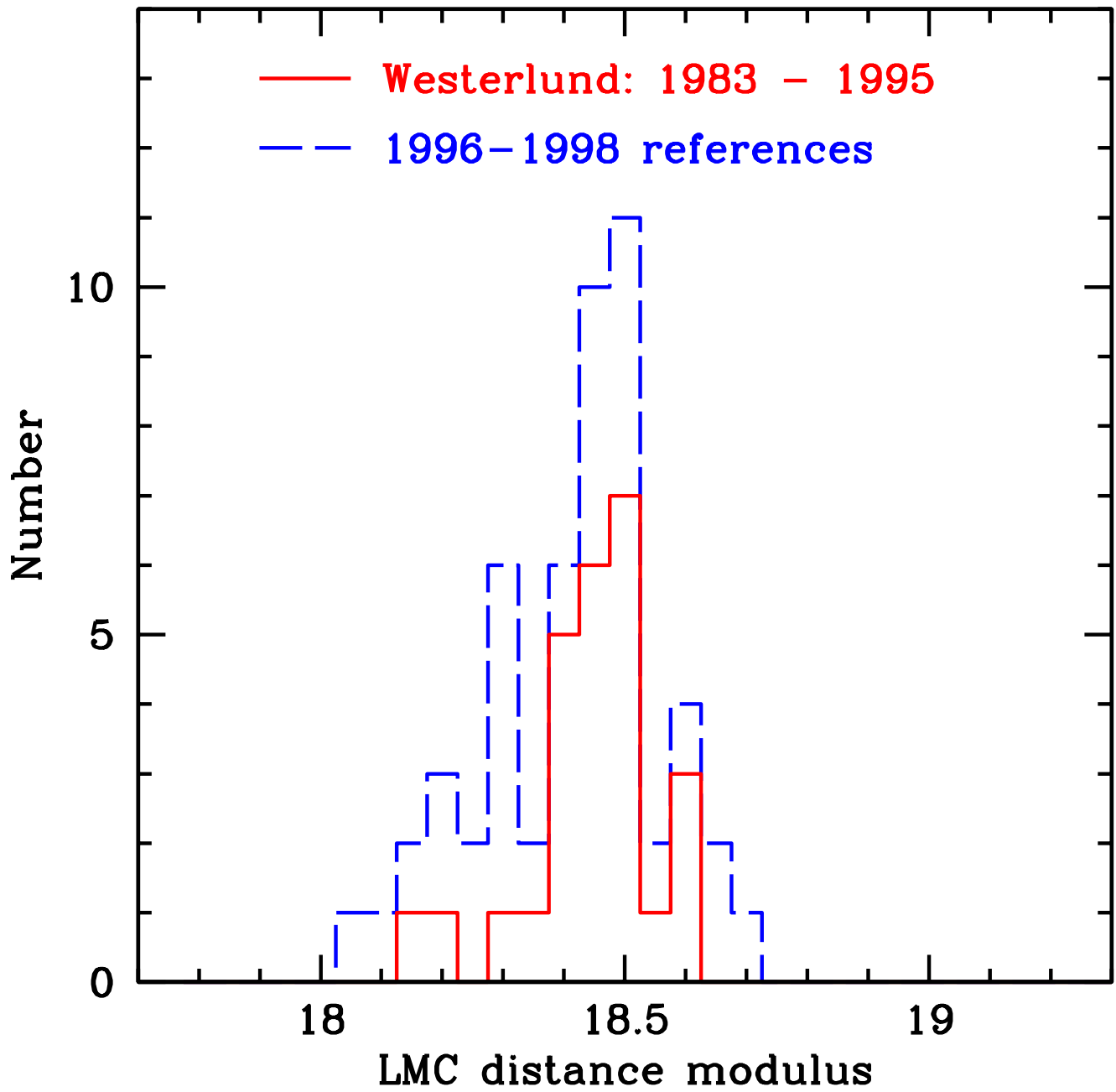
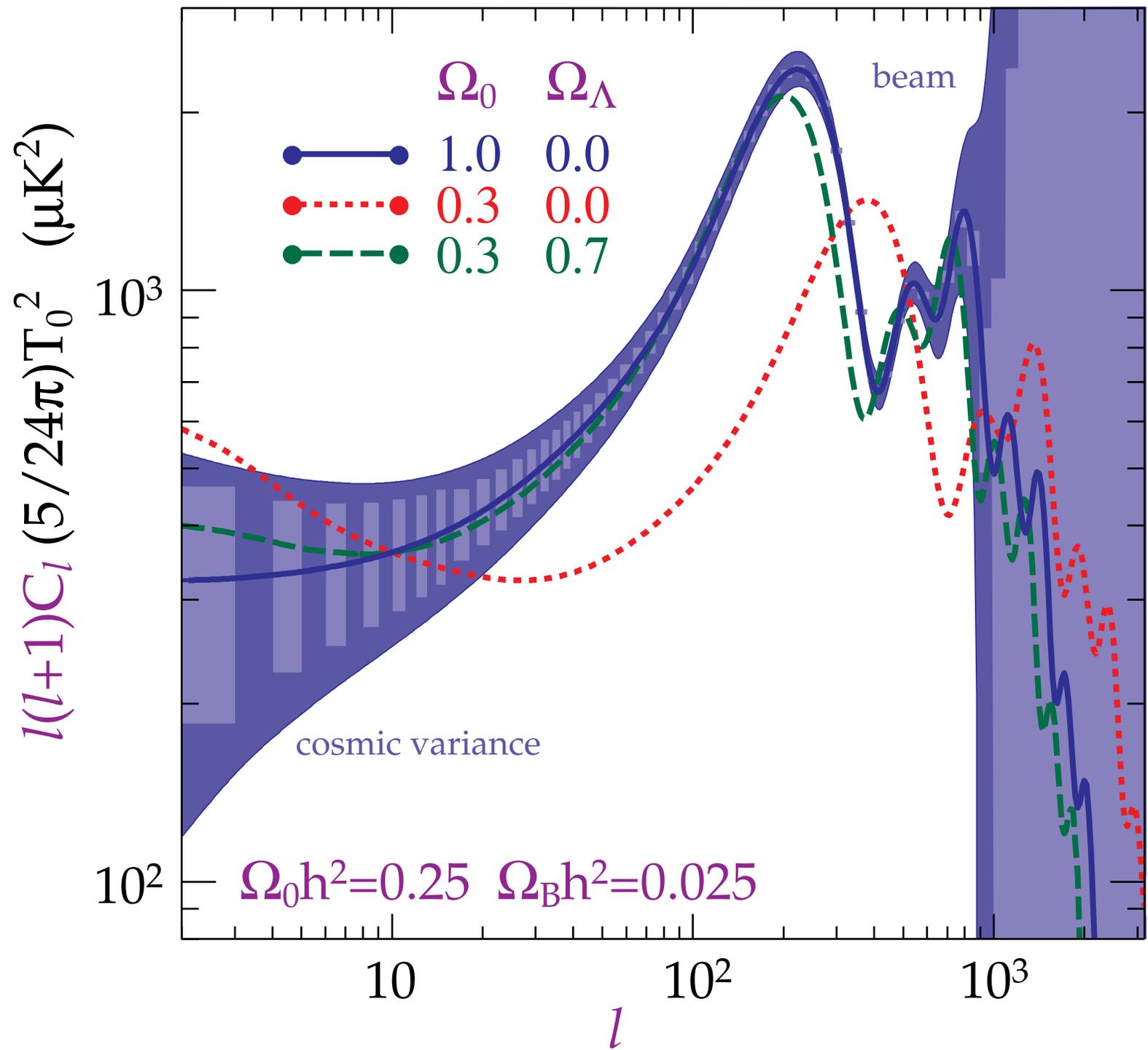
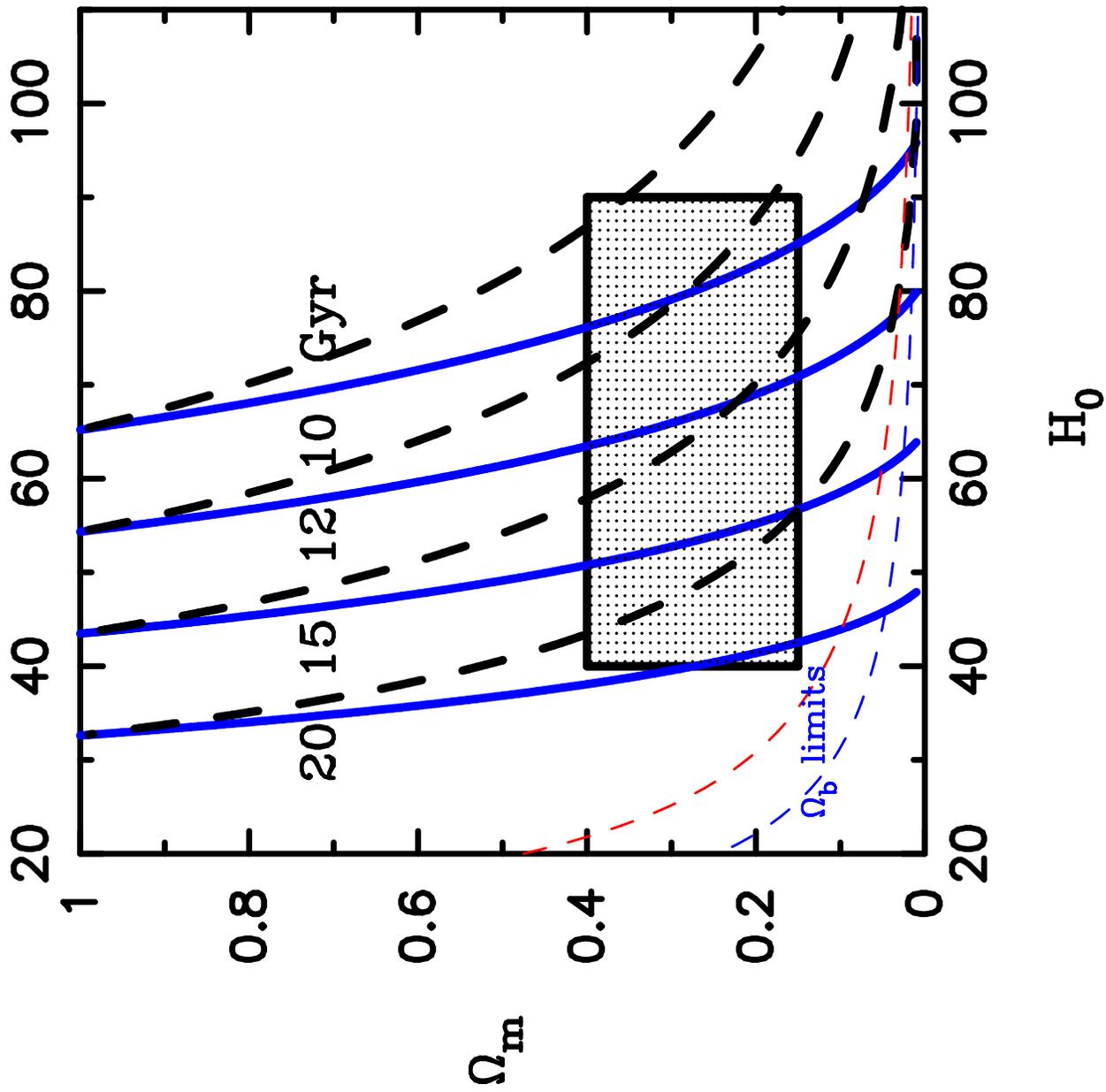
\end{document}